\documentclass[12pt,preprint]{aastex}
\usepackage{graphicx}
\usepackage{url}
\newcommand{\beq}{\begin{equation}}
\newcommand{\eeq}{\end{equation}}
\newcommand{\beqa}{\begin{eqnarray}}
\newcommand{\eeqa}{\end{eqnarray}}

\def\la{\lower.5ex\hbox{$\; \buildrel < \over \sim \;$}}
\def\ga{\lower.5ex\hbox{$\; \buildrel > \over \sim \;$}}

\begin{document}
\title{Dynamics of the Local Group: the Outer  \\ Galactic Globular Star Clusters}
\author{P.~J.~E. Peebles}  
\affil{Joseph Henry Laboratories, Princeton University, Princeton, NJ 08544}
\begin{abstract}
A model for the mass in and around the Local Group previously used to fit redshifts of dwarf galaxies to their distances between 50~kpc and 2.6~Mpc under the condition of small and growing primeval departures from homogeneity is shown to allow fits to distances and redshifts of twelve galactic globular clusters at galactocentric distances greater than 30\,kpc. The solutions also fit three sets of measured globular cluster proper motions and the orientation of one observation of tidal tails. In some solutions these outer globular clusters have circled the Milky Way several times, losing information about their initial conditions. In other trajectories globular clusters are approaching the Milky Way for the first time from formation in mass concentrations modest enough to have small internal velocities and initially moving away from the proto--Milky Way galaxy at close to the general rate of expansion of the universe. \end{abstract}
\maketitle

\section{Introduction}\label{sec:1}

This analysis of the dynamics of the outer galactic globular star clusters (GGCs) complements the measure of the mass distribution in the Milky Way (MW) galaxy from the distribution of GGCs in their single-particle phase space (Eadie, Springford, and Harris 2017a,b). The phase space distribution is well sampled at galactocentric distances less than 30\,kpc, but the sample is sparse further out. The complementary approach presented here computes trajectories of GGCs at galactocentric distances greater than 30\,kpc moving as massless test particles in the evolving mass distribution used in analysis of the dynamics of dwarf galaxies at distances greater than 50\,kpc (Peebles 2017; hereinafter P17). A comparison of computed trajectories of GGCs and dwarf galaxies is aided by adding six dwarf galaxies at distances from 30\,kpc to the 50\,kpc limit in P17. 

The analysis uses the Numerical Action Method (NAM, in the variant presented in the Appendix in Peebles, Tully, and Shaya 2011) to find trajectories under mixed boundary conditions. The final condition is that the object ends up at its observed present distance, within the measurement uncertainty. The initial condition is that the object has small and growing peculiar motion at high redshift produced by the peculiar gravitational acceleration of the early close to homogeneous mass distribution. The check is that solution has an acceptable redshift. When applied to the motion of a massive proto-galaxy the initial  condition is meant to approximate the motion of the center of mass of the matter that is merging to form the galaxy. It would also apply to ``quiet'' or ``primeval'' formation of globular clusters, or dwarf galaxies,  in small mass concentrations with initial peculiar velocities close to linear perturbation theory.  An early variant of this  picture is the Peebles and Dicke (1968) proposal that globular clusters formed at the primeval baryon Jeans mass during the early onset of nonlinear departures from homogeneity. In a recently discussed variant GGCs formed in dark matter halos of dwarf galaxies that may later have been tidally disrupted, leaving apparently isolated GGCs (Zaritsky, Crnojevi{\'c}, and Sand 2016 and references therein). In this picture the quiet initial conditions would be a good approximation if the dwarf galaxy internal velocities were subdominant to the peculiar velocities $\sim100$~km~s$^{-1}$ driven by the early large-scale mass distribution. The quiet picture would not apply to globular clusters or dwarf galaxies that dissipatively or tidally formed or were cast off by gravitational field fluctuations during formation of massive proto-galaxies, as in  the Fall and Rees (1977) picture for globular clusters. The issues constraining theories of globular cluster formation are far richer now (as reviewed by Brodie and Strader 2006; Renzini 2017; and Kim,  Ma, Grudi{\'c}, et al. 2017). But there remains a fundamental distinction between formation of globular clusters in close to quiet primeval conditions, as assumed in the NAM solutions, or in the large internal velocities and dissipative conditions in massive proto-galaxies. 

NAM solutions may be expected to better fit measured positions and motions of objects that formed under  quiet primeval conditions rather than the active conditions in massive proto-galaxies. However, there is the complication that the NAM mixed boundary conditions allow multiple solutions, one of which may accidentally fit the observed distance and redshift with the initial conditions of the quiet picture even though the object did not form that way. Also hindering distinction between modes of formation is the fact that the mass model necessarily is at best a crude approximation. An illustration of a possible flaw in the mass model is the systematic underestimate of redshifts and/or overestimate of distances of the six dwarfs in the NGC\,3109 association and the two in the DDO\,210 association, at measured distances 1.1 to 1.6~Mpc (P17 Table 4). It is the consistency of the P17 fits to measured redshifts and distances of 49 dwarfs at 50~kpc to 1~Mpc that  argues for the quiet picture for formation of these dwarf galaxies, and for reasonable adequacy of the model for the mass distribution and evolution at distances less than about 1~Mpc.

Section~\ref{analysis} presents the extension of this test of the mass model and initial conditions to the measurements of twelve GGCs at galactocentric distances greater than 30~kpc and six dwarf galaxies at distances between 30 and 50~kpc. These objects, being closer to MW, may be more likely to have completed several passages around the galaxy, allowing imperfections of the mass model to erase signatures of initial conditions. Section~\ref{sec:degeneracies} addresses this by comparing trajectories in multiple solutions that fit the measurements reasonably well. Degeneracy might be reduced by the observed orientations of tidal tails. This is briefly considered in Section~\ref{sec:tails}. A concluding assessment what has been learned is offered in Section~\ref{sec:discussion}.

\section{Analysis}\label{analysis}

Following P17, the GGCs and dwarf galaxies considered here are taken to move as massless test particles, or tracers, under the gravitational attraction of  twelve massive actors moving in accordance with their gravitational interactions, and neglecting stellar dynamical drag. In P17 the parameters of the mass model were adjusted to get the best fit to all the redshifts, distances, and proper motions, along with the relative masses indicated by the luminosities of the massive actors. The final mass model from P17 with halo shape parameter $\alpha=6$ is used here without further adjustment. 

NAM yields solutions to the equation of motion with primeval initial conditions by relaxing to solutions at stationary points of the action starting from trial trajectories. For the relatively close objects considered here trial paths were obtained by integrating back in time from trial distances randomly drawn from a uniform distribution within 20\% of the catalog value, and trial redshifts drawn from a uniform distribution within 30~km~s$^{-1}$ of catalog. When available, proper motions for trial trajectories were  drawn from uniform distributions centered on the catalog values within 2.5 times the measurement uncertainties, and where not available randomly drawn transverse motions in orthogonal directions in the range $\pm400$~km~s$^{-1}$. 

Fits of the solutions to the data are ranked by a $\chi^2$ measure, as in P17. It should be understood that although $\chi^2$ offers a convenient way to rank goodness of fits of solutions to measurements its value has even less significance here than in P17 because the measurement uncertainties may be expected to be subdominant to errors introduced by the schematic model for the distribution of mass in and around MW. (The remarkable exception is the GGC Palomar\,2, whose listed redshift uncertainty is $\pm 57$\,km\,s$^{-1}$.) In the computation of $\chi^2$ the nominal standard deviations in redshift are the minimum of 5~km~s$^{-1}$ or the catalog uncertainty, nominal standard deviations in distances are 5~\%, and the catalog uncertainties of proper motion measurements are treated as standard deviations. The distance uncertainty seems to be a reasonable approximation to the state of the art. The adopted redshift uncertainty is far larger than some measurements; it is meant to acknowledge that the best possible result of application of the approximate mass model is an approximation to the true trajectory.  

\begin{table}[htpb]
 \centering
\begin{tabular}{lrrrrrrrr}
\multicolumn{8}{c}{Table 1: GGC and nearby dwarf galaxy distances and redshifts} \\
\noalign{\medskip}
\tableline\tableline\noalign{\smallskip}
 & \multicolumn{2}{c} {catalog distance} & model &&  \multicolumn{2}{c} {catalog redshift}  & model &$v_i$ \\
  \cline{2-3} \cline{6-7} \\
\noalign{\vspace{-5mm}}
 &   GC  &  HC &  HC  & & GC  & HC & HC \\
\noalign{\vspace{1mm}}
\tableline
\noalign{\vspace{1mm}}
*Bootes III       &     45.8 &$     46.8 \pm      2.3 $&     44.7 &&$   240 $&$  197 \pm    5 $&$   186 $&    68\\
Coma Berenices    &     44.8 &$     43.7 \pm      2.2 $&     38.0 &&$    82 $&$   98 \pm    5 $&$    91 $&   138\\
Willman 1         &     42.6 &$     38.0 \pm      1.9 $&     34.1 &&$    34 $&$  -12 \pm    5 $&$   -16 $&    84\\
Segue II          &     40.5 &$     34.7 \pm      1.7 $&     37.7 &&$    40 $&$  -39 \pm    5 $&$    -2 $&    70\\
Bootes II         &     39.5 &$     41.7 \pm      2.1 $&     41.4 &&$  -115 $&$ -117 \pm    5 $&$  -117 $&    11\\
Ursa Major II     &     37.7 &$     31.6 \pm      1.6 $&     32.6 &&$   -35 $&$ -116 \pm    5 $&$  -113 $&    87\\
Crater/Laevens\,I   &    147.8 &$    148.0 \pm      7.4 $&    151.1 &&$     0 $&$  148 \pm    5 $&$   146 $&    73\\
AM 1              &    124.6 &$    123.3 \pm      6.3 $&    123.2 &&$   -39 $&$  116 \pm   20 $&$   151 $&    61\\
Pal 4             &    111.3 &$    108.7 \pm      5.4 $&    104.1 &&$    50 $&$   74 \pm    5 $&$    78 $&    39\\
Pal 3             &     95.8 &$     92.5 \pm      4.6 $&     85.1 &&$   -64 $&$   83 \pm    8 $&$    66 $&    33\\
Eridanus          &     95.0 &$     90.1 \pm      4.5 $&     91.3 &&$  -141 $&$  -23 \pm    5 $&$   -24 $&   113\\
NGC 2419          &     89.9 &$     82.6 \pm      4.1 $&     87.1 &&$   -28 $&$  -20 \pm    5 $&$   -19 $&    51\\
Pal 14            &     71.6 &$     76.5 \pm      3.8 $&     76.1 &&$   166 $&$   72 \pm    5 $&$    75 $&   109\\
Pyxis             &     41.4 &$     39.4 \pm      2.0 $&     39.6 &&$  -191 $&$   34 \pm    5 $&$    25 $&    86\\
NGC 7006          &     38.6 &$     41.2 \pm      2.1 $&     41.6 &&$  -187 $&$ -384 \pm    5 $&$  -388 $&   133\\
Pal 15            &     38.4 &$     45.1 \pm      2.3 $&     45.5 &&$   149 $&$   68 \pm    5 $&$    69 $&    91\\
Pal 2             &     35.0 &$     27.2 \pm      1.4 $&     27.0 &&$  -107 $&$ -133 \pm   57 $&$  -156 $&    88\\
Whiting 1         &     34.6 &$     30.1 \pm      1.5 $&     30.1 &&$  -106 $&$ -130 \pm    5 $&$  -132 $&    70\\
\tableline
\noalign{\smallskip}
\multicolumn{4}{l}{Units: kpc and km s$^{-1}$} \\
\end{tabular}
\end{table}

\subsection{Measured and Modeled Redshifts, Distances, and Proper Motions}\label{sec:redshiftdistance}
Table 1 lists measured redshifts and distances from McConnachie (2012; 2015) for the six dwarf galaxies at galactocentric distances between 30~kpc and the 50~kpc limit in P17. They are followed by twelve GGCs with galactocentric distances greater than 30~kpc. Crater-Laevens\,I is not a recognized and generally accepted GGC, but recent detailed studies make a strong case for it (Voggel, Hilker, Baumgardt, et al. 2016; Weisz, Koposov, Dolphin, et al.\ 2016). The relatively large distance makes this object particularly interesting. Its angular position is from Laevens, Martin, Sesar, et al. (2014), and the redshift and distance are from Voggel, Hilker, Baumgardt, et al. (2016). The redshifts and distances of the other twelve GGCs are from Eadie, Springford, and Harris (2017a), with angular positions from Harris (2010). The entries are listed in decreasing order of the catalog galactocentric (GC) distances in the second column derived from the catalog heliocentric (HC) distances in the third column with MW distance 8\,kpc. Galactocentric distances are relevant for these relatively close objects because the multiplicity of solutions that pass close to MW several times tends to increase with decreasing galactocentric distance. The measured heliocentric distances are to be compared to the heliocentric distances in the fourth column from the NAM solutions that best fit the measurements. The GC redshift in the fifth column is the component of the galactocentric velocity of the object in the direction of the heliocentric angular position. This is a reasonable approximation to the galactocentric radial velocity. The catalog heliocentric redshifts in the sixth column are to be compared to the model heliocentric redshifts in the seventh column. The last column lists the physical peculiar velocity $v_i$ of each object at redshift $1+z=10$ in the best-fitting solution. 

\begin{table}[ht]
\centering
\begin{tabular}{lrrrrr}
\multicolumn{6}{c}{Table 2: Proper Motions$^{\rm a}$}\\
\noalign{\medskip}
\tableline\tableline\noalign{\smallskip}
  & \multicolumn{2}{c}{$\mu_\alpha$} & \  &  \multicolumn{2}{c}{$\mu_\delta$}  \\
   \noalign{\smallskip}
  \cline{2 - 3}  \cline{5 - 6}
 &  measured\hspace{2mm}  & model &\ & measured\hspace{2mm} & model  \\
\tableline
 \noalign{\smallskip}
   Pal 3              &$     0.33 \pm     0.23 $&$     0.57 $&\ &$     0.30 \pm     0.31 $&$     0.30 $ \\
  Pyxis              &$     1.09 \pm     0.31 $&$     1.85 $&\ &$     0.68 \pm     0.29 $&$     1.09 $ \\
  NGC 7006           &$    -0.96 \pm     0.35 $&$    -1.60 $&\ &$    -1.14 \pm     0.40 $&$    -0.61 $ \\
\tableline
 \noalign{\smallskip}
\multicolumn{6}{l}{$^{\rm a}$milli arc sec y$^{-1}$}\\
\end{tabular}
\end{table}

Table 2 lists measured GGC proper motions (from the tabulation in Eadie et al. 2017a for Palomar\,3 and NGC\,7006, and Fritz, Linden,  Zivick, et al.\ 2017 for Pyxis).  Columns three and five show the model values in the solutions with the least nominal $\chi^2$. 

Since the $\chi^2$ measures of fit have doubtful significance they are not listed in Tables~1 and~2. I instead take note of the largest apparent discrepancies. The dwarf galaxy Segue\,II has the most serious difference between catalog and model redshifts: catalog value $-39$~km~s$^{-1}$, model value $-2$~km~s$^{-1}$. All the other differences between model and measured redshifts are within two nominal standard deviations. (The GGC AM\,1 has model redshift $35$~km~s$^{-1}$ larger than catalog, but its catalog uncertainty is $20$~km~s$^{-1}$.)  The two most serious distance discrepancies are are for the dwarf Coma Berenices, with model value 38~kpc and measured 44~kpc, and the dwarf Willman\,1, with model distance 34~kpc and measured 38~kpc.  All other model distances are close to or within 5\% of catalog.  The model proper motion of Pyxis in the direction of increasing right ascension is larger than the measurement by 2.4 times the measurement uncertainty. All other components are well within twice the measurement uncertainty. 

\begin{figure}[ht]
\begin{center}
\includegraphics[angle=0,width=4.5in]{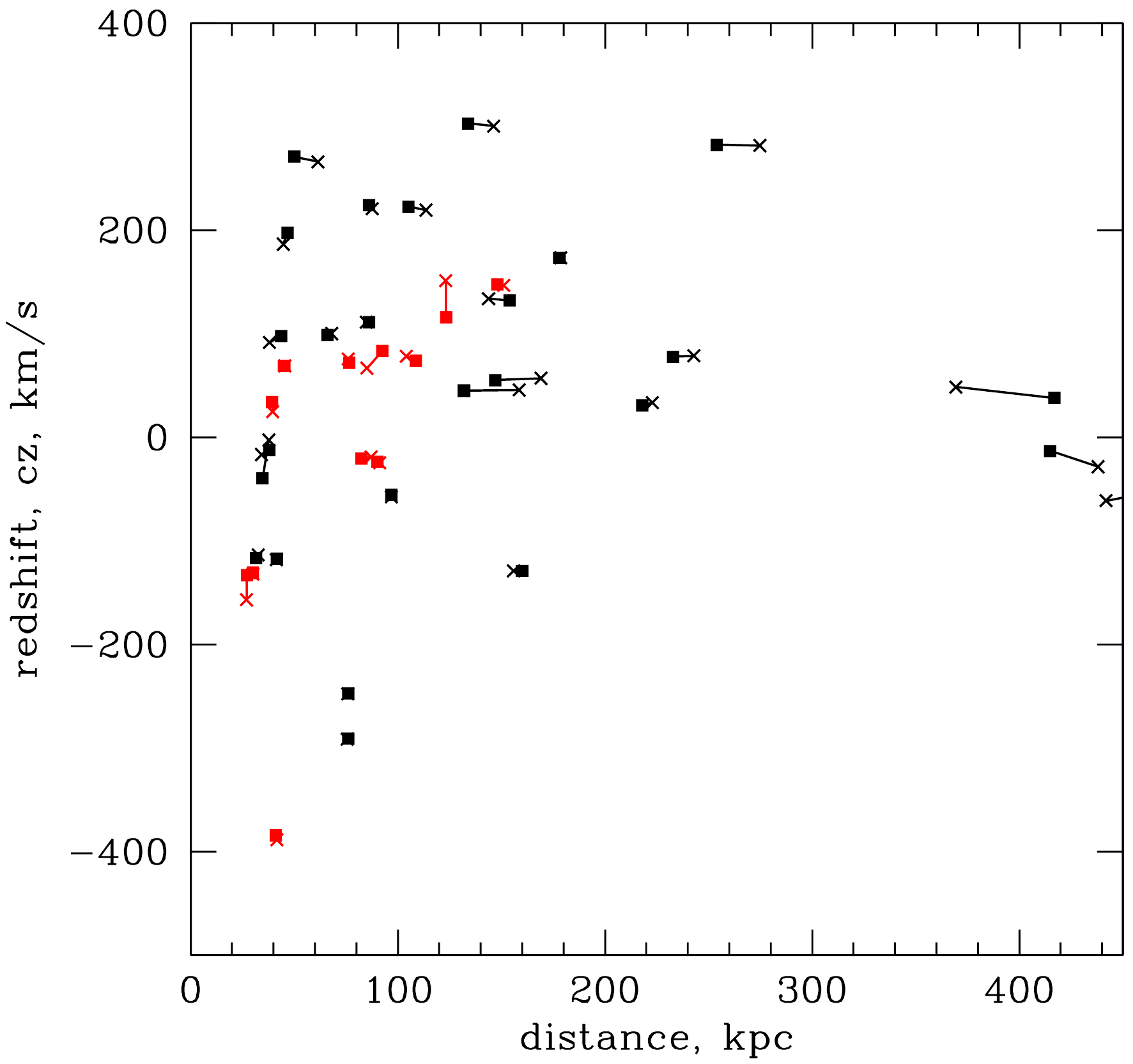} 
\caption{\small Catalog distances and redshifts plotted as squares are connected by lines to model values plotted as crosses, in black for dwarf galaxies and red for galactic globular clusters.}
\label{Fig:cz}
\end{center}
\end{figure}

The model results may be assessed also by the visual comparison of model and measured redshifts and distances in Figure~1. The black symbols represent dwarf galaxies, the red GGCs. Model distances and redshifts plotted as crosses are connected by lines to the catalog values plotted as filled squares. The fit of model to measurements for the dwarfs at distances greater than 50\,kpc benefitted from the adjustment of the mass model parameters in P17 that minimize an overall $\chi^2$ measure of fit (though it must be noted that there are so many parameters in P17 that a poor value of $\chi^2$ for one object has modest effect on the mass model). The nearer six dwarfs added for this analysis fit the measurements about equally well with no further adjustment of the mass model. The quality of fits for the twelve outer GGCs shown in red is not noticeably different from the fits to dwarf galaxies at similar distances. This discussion continues in Section~\ref {sec:discussion}.

\subsection{Degenerate Solutions}\label{sec:degeneracies}
The multiple solutions allowed by the NAM mixed boundary conditions tend to be particularly abundant and difficult to interpret when the tracer is close enough to MW to allow completion of several orbits in a Hubble time. The situation is illustrated by the examples presented here. 

\begin{figure}[ht]
\begin{center}
\includegraphics[angle=0,width=4.5in]{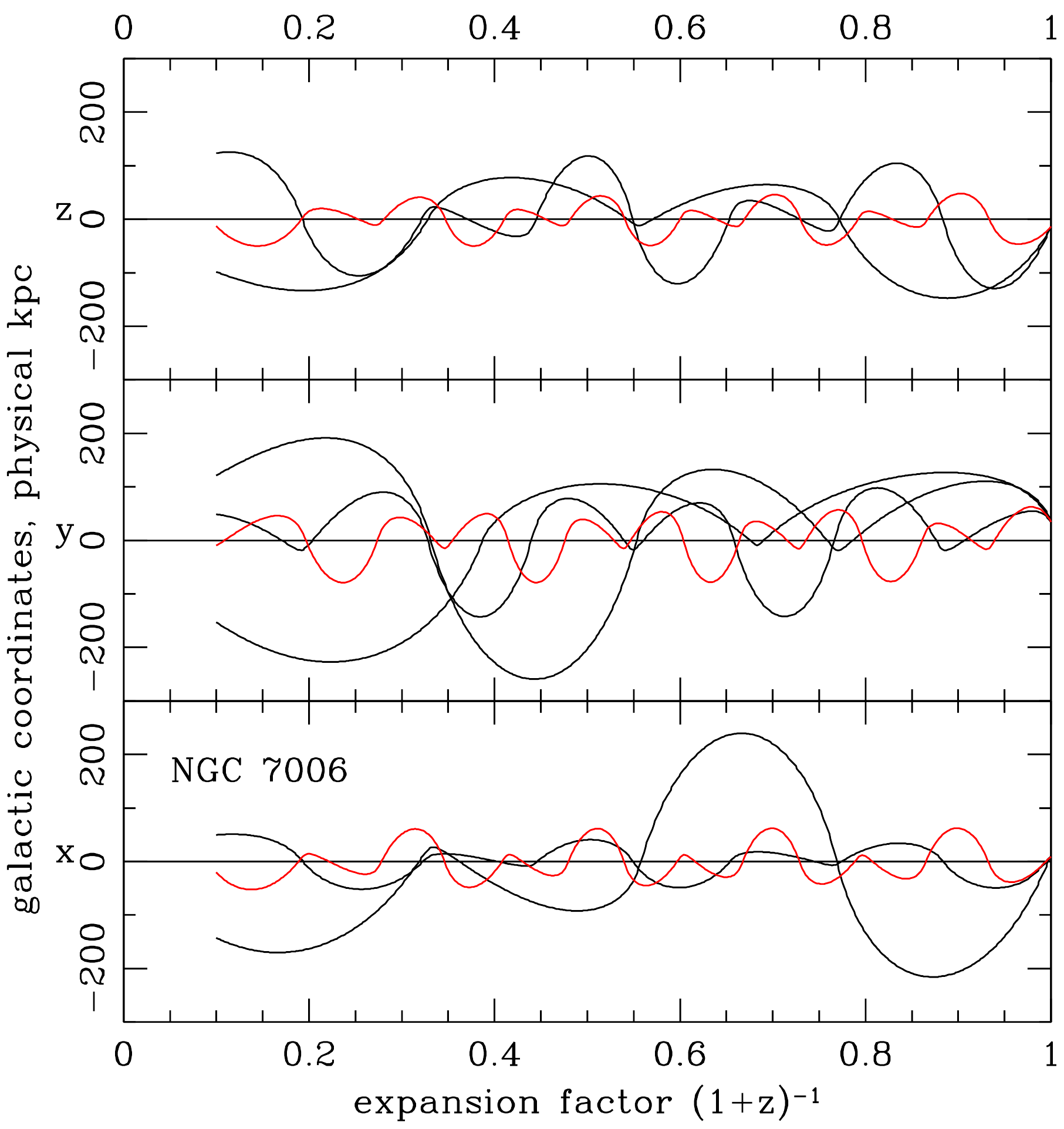} 
\caption{\small Solutions for the motion of the globular cluster NGC\,7006 relative to MW in physical galactic coordinates as functions of the expansion factor, where $z$ is the redshift. The two black curves satisfy primeval initial conditions. The red curve is computed back in time from central values of the measured position and velocity.}
\label{Fig:N7006}
\end{center}
\end{figure}
 
\begin{figure}[h]
\begin{center}
\includegraphics[angle=0,width=4.5in]{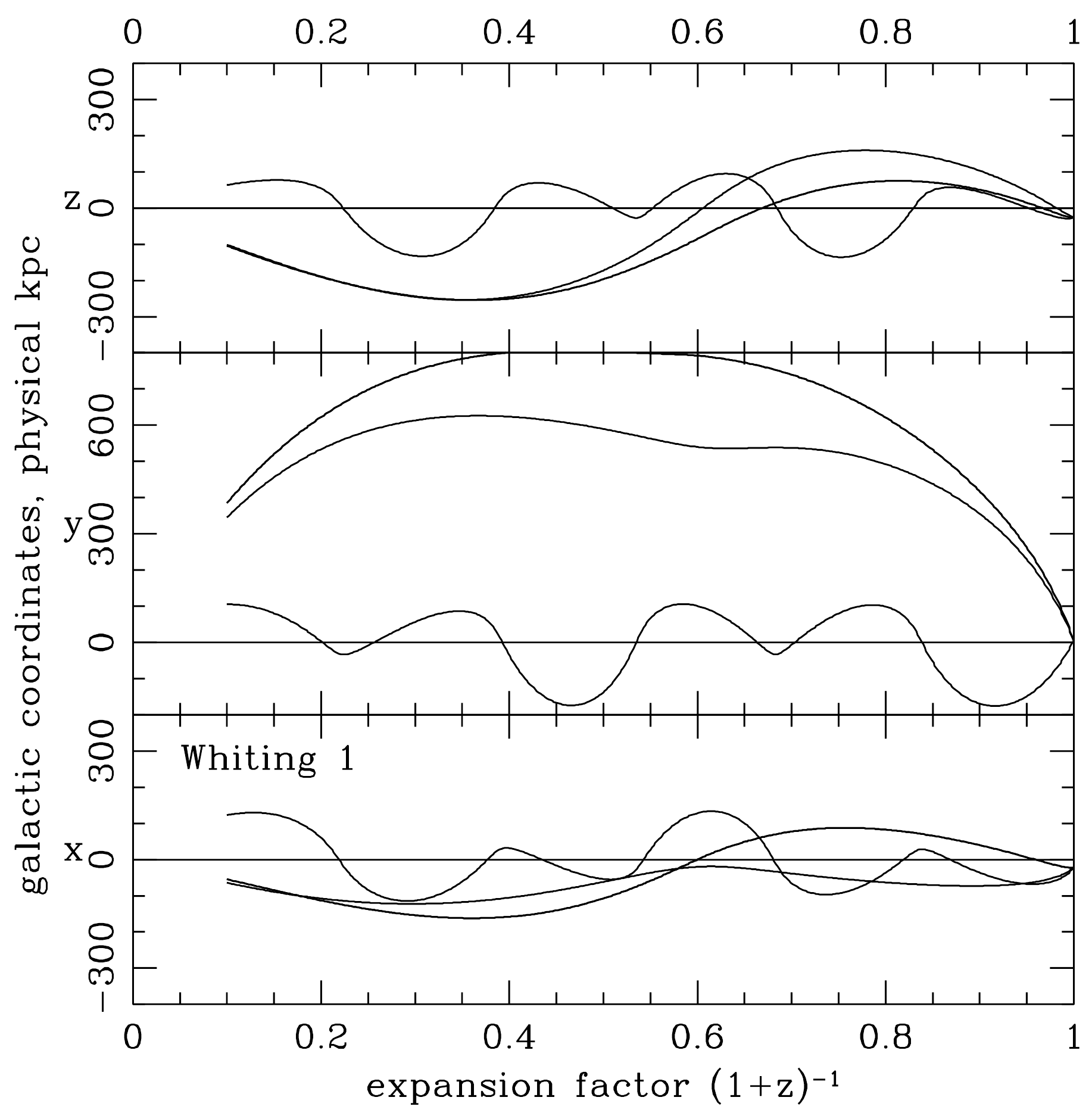} 
\caption{\small Solutions for the trajectory of the globular cluster Whiting\,1, plotted as in Fig.~\ref{Fig:N7006}.}
\label{Fig:Whiting}
\end{center}
\end{figure}

Figure~\ref{Fig:N7006} shows solutions to the equations of motion for the GGC NGC\,7006 fitted to its measured  proper motion, distance, and redshift. The right-hand coordinate system is galactic with the x-axis pointing in the direction $\ell = 0 = b$; the lengths are physical; and the origin is fixed to MW. The solutions follow the last factor of ten expansion of the universe. The two black curves in Figure~\ref{Fig:N7006} satisfy the quiet primeval initial conditions. The model redshifts, distances, and proper motions all are within two nominal standard deviations of catalog. The solution shown as the red curve is computed back in time from the central values of the catalog distance, redshift, and proper motion; the conditions at the high redshift end of this red curve are not constrained. This red trajectory has circled MW three times, passing within 20 kpc of MW twelve times (with close to three perigalacticons per orbit). We must expect that the imperfections in the MW mass model, and the neglect of dynamical drag, have seriously distorted the computed path of NGC\,7006. That is, details of the early position and velocity in this red solution are meaningless, apart from the fact that the trajectory stays relatively close to MW. We must similarly expect that, although the initial conditions of the black curves are consistent with a  primeval origin, these trajectories also have passed close to MW often enough to have made the initial conditions meaningless. That is, these solutions indicate NGC\,7006 has circled MW several times, and likely formed somewhere between 50 and 200 physical kpc from the proto-MW, but they offer no evidence for or against the picture of quiet formation outside the major pieces of the proto-MW. 

The acceptable solutions in Figure~\ref{Fig:Whiting} for Whiting\,1, the globular cluster closest to MW in the sample considered here, illustrate another aspect of the situation. The three solutions satisfy primeval initial conditions, and all three fit the measured redshifts and distances close to or within the nominal standard deviations.  Two of these  solutions have Whiting\,1 approaching MW for the first time, from maximum physical distance of about 700~kpc. These paths are simple enough, and well enough away from MW apart from the recent approach, that their initial conditions seem to be reasonably well based within the parameters of the mass model. But the third solution has Whiting\,1 completing several passages of MW. This solution likely has strayed far from the initial conditions, whether primeval or active, that would obtain in a more realistic mass model. This cluster could be falling into MW for the first time, or it could have completed several orbits of MW. 

The dwarf galaxy Ursa Major\,II has about the same galactocentric distance as Whiting\,1, and the three solutions with reasonable fits to the measurements  show a similar situation: one has Ursa Major\,II approaching MW for the first time, two have Ursa Major\,II circling MW closely enough to lose track of initial conditions.

\begin{figure}[ht]
\begin{center}
\includegraphics[angle=0,width=4.5in]{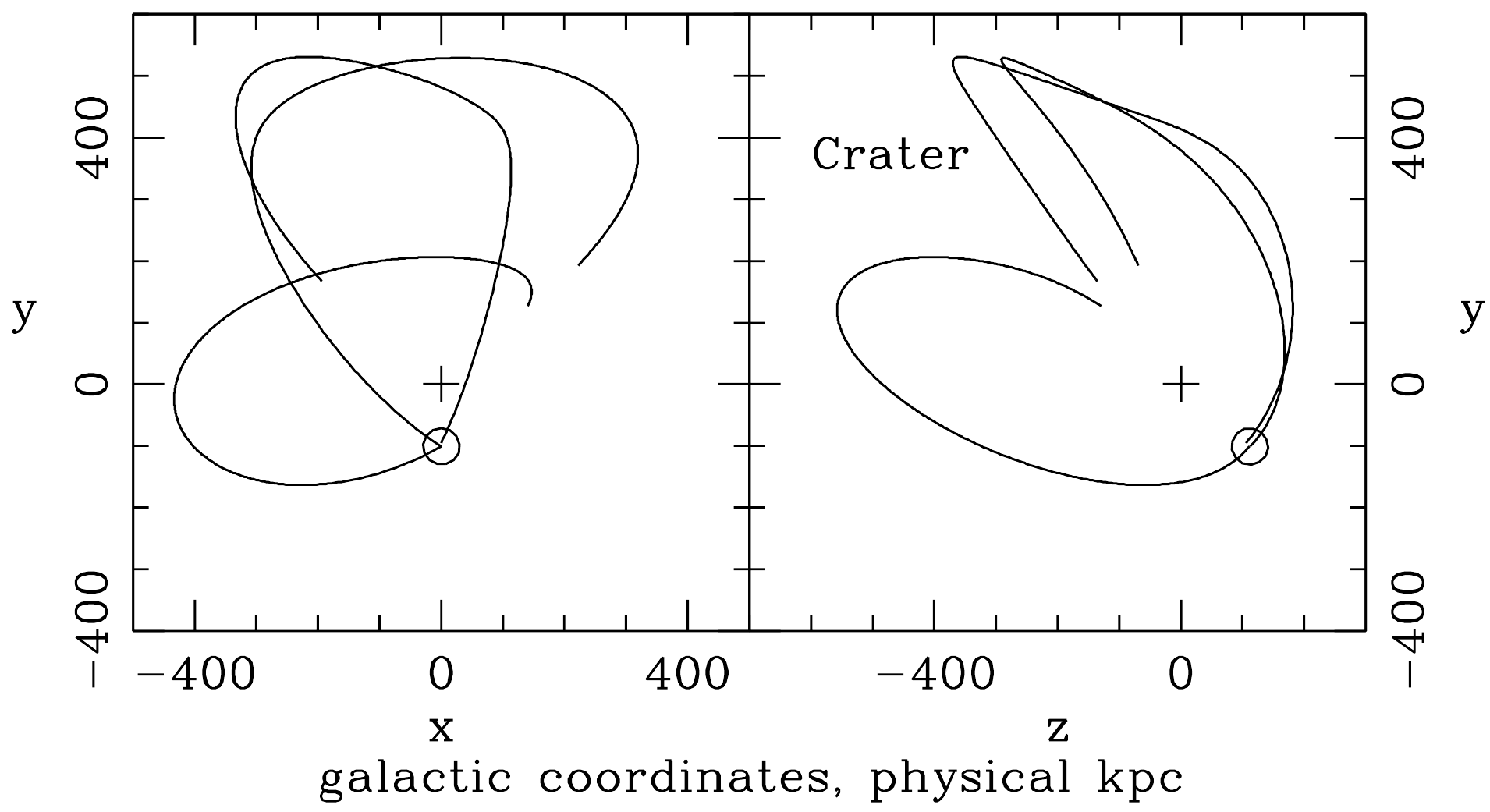} 
\caption{\small The best-fitting solutions for Crater/Laevens\,I. MW is at the origin at the plus sign. The circle marks the present position of Crater/Laevens\,I.}
\label{Fig:Crater}
\end{center}
\end{figure}

The three solutions with cosmological initial conditions and acceptable fits to the measured redshift and distance of the most distant GGC in this sample, Crater/Laevens\,I, are shown in Figure~\ref{Fig:Crater}. As in Figures~\ref{Fig:N7006} and~\ref{Fig:Whiting} the galactic coordinates are physical, but here the trajectories are plotted in orthogonal projections with the origin of coordinates at the plus sign fixed to the path of MW.  The circle marks the present position of Crater/Laevens\,I. All three solutions fit the measured distance and redshift within the nominal standard deviations, and all have this object approaching MW for the first time, from quite different directions.

 \begin{figure}[ht]
\begin{center}
\includegraphics[angle=0,width=4.5in]{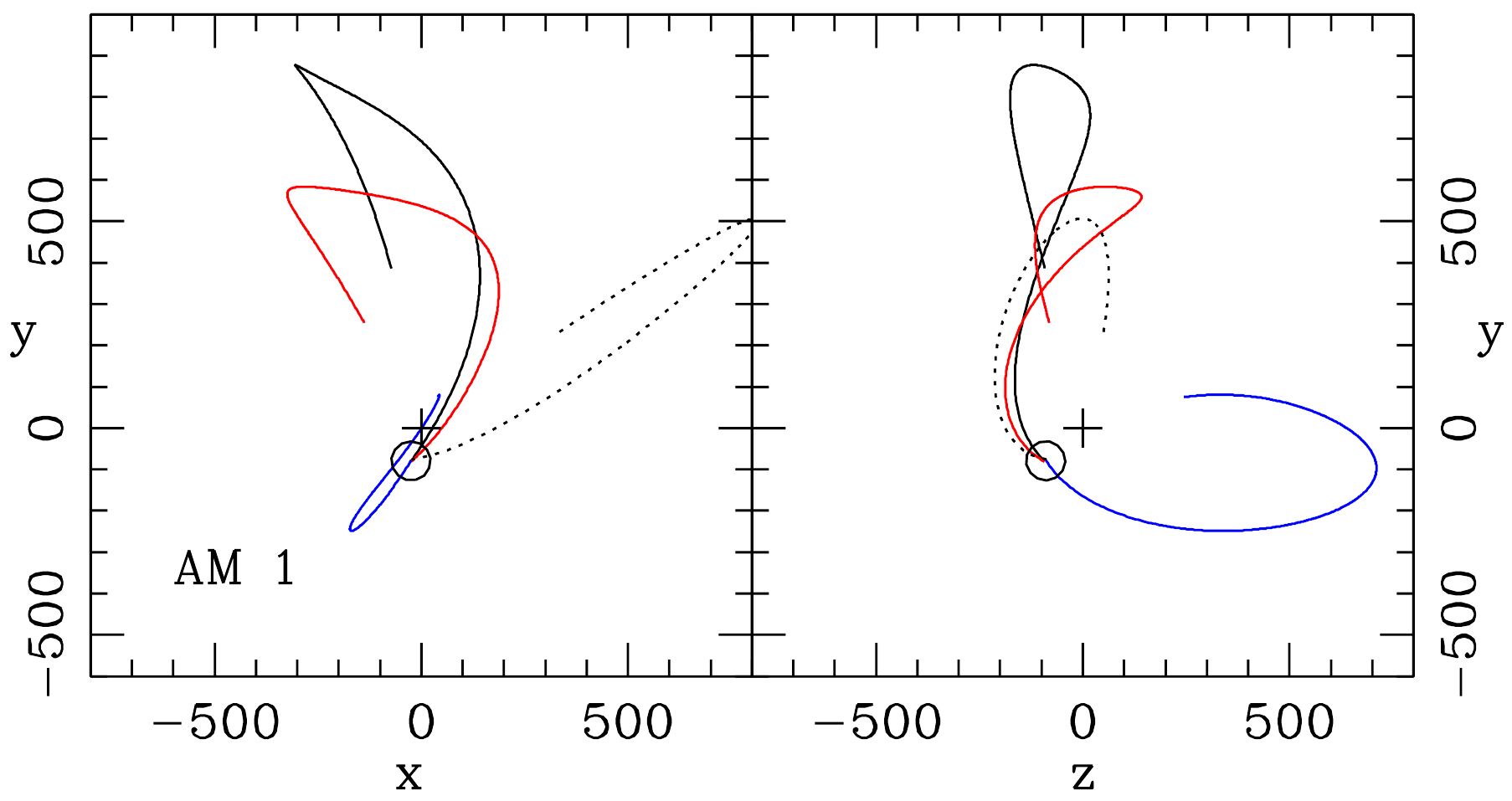} 
\caption{\small Solutions for the trajectory of the globular cluster AM\,1,  plotted as in Figure \ref{Fig:Crater}. The solutions that fit the catalog redshift and distance about equally well are plotted as the solid black, red, and blue curves. The next best, and unpromising, fit is shown as the dotted black curve.}
\label{Fig:AM}
\end{center}
\end{figure}

The acceptable solutions with cosmological initial conditions for the next most distant GGC, AM\,1, are shown in Figure~\ref{Fig:AM}.  The present distances to AM\,1 in all four solutions are at or within one nominal standard deviation. The solution that best fits the redshift is plotted as the solid black curve, the next best is the red curve, and the next best the blue curve. The redshifts of these three solutions are within two nominal standard deviations of catalog. The dashed curve has redshift 85~km~s$^{-1}$ below catalog, which looks quite unpromising, even though the catalog uncertainty is 20~km~s$^{-1}$. All four solutions have AM\,1 approaching MW for the first time, from maximum distances on the order of 800~kpc, and from initial conditions that seem to be reasonably secure within the model.

\begin{figure}[ht]
\begin{center}
\includegraphics[angle=0,width=4.5in]{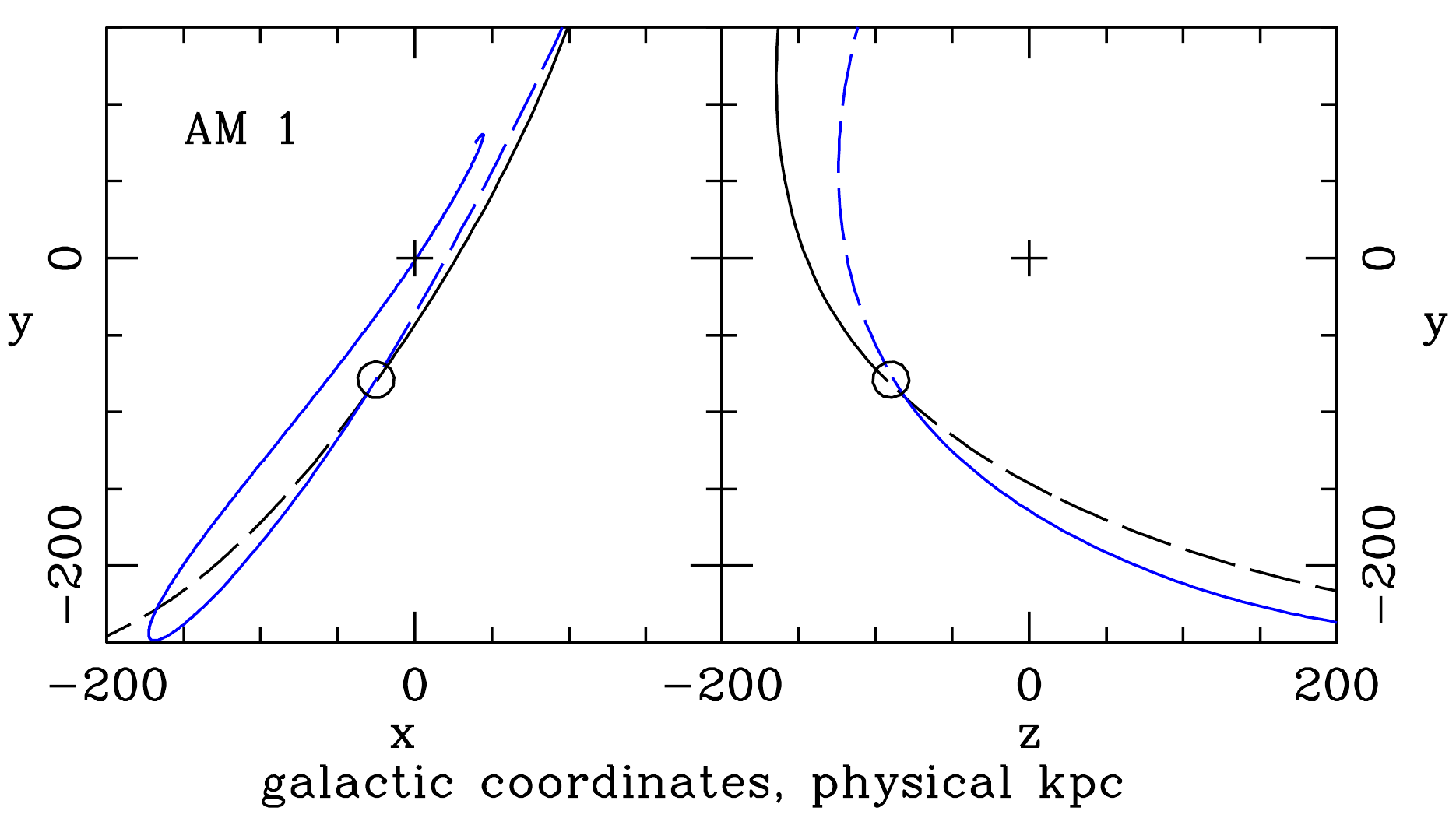} 
\caption{\small Dashed curves are trajectories of AM\,1 computed forward in time from $z=0$ with initial conditions from the solutions plotted as solid curves here and in the same color code as in Figure~\ref{Fig:AM}.}
\label{Fig:AMfwd}
\end{center}
\end{figure}

The direction of approach of AM\,1 to MW in the solution plotted in blue in Figure~\ref{Fig:AM} is almost directly opposite the solution plotted in black. This is shown in more detail in Figure~\ref{Fig:AMfwd}. The dashed curves show numerical integrations forward in time from positions and velocities at redshift $z = 0$ taken from the solutions plotted as the solid curves in Figures~\ref{Fig:AM} and~\ref{Fig:AMfwd}. The integration forward from the present takes account only of the gravitational acceleration of MW, and the small effect of the cosmological constant $\Lambda$, which is a reasonable approximation for the range of distances shown in Figure~\ref{Fig:AMfwd}. The dashed curves run off scale in the figure when the universe has expanded by a factor $\sim 1.06$ from the present. At seriously larger expansion factors the  integrations forward in time are not to be trusted because they ignore the approaching M\,31. The near time-reversal symmetry seen in Figure~\ref{Fig:AMfwd} might be understood by considering the trajectory of a tracer particle from quiet primeval conditions when a spherical MW is completely isolated from other mass. If the orbital energy is negative then at small early separations the tracer moves away from MW at close to the Friedman equation with zero space curvature, and it later falls back radially through MW and moves away in a time-reversal of the approach, apart from the small effect of $\Lambda$. If the trajectory is perturbed by adding angular momentum relative to MW then the tracer at perigalacticon, observed from the center of MW, would have zero redshift. This would allow two trajectories, running in the directions of increasing and decreasing time. The redshift of AM\,1 relative to MW is fairly small, $-40$~km~s$^{-1}$. Perhaps the breaking of time reversal invariance by the evolving exterior mass distribution has been modest enough to allow the impression of approximate time-reversal symmetry in Figure~~\ref{Fig:AMfwd}.  It might be noted that the trajectory of Crater/Laevens\,I that approaches MW from below the plane of MW is approaching roughly antiparallel to one of the solutions approaching from above the plane. Maybe this is consistent with the small galactocentric radial velocity of this cluster. 

\begin{figure}[ht]
\begin{center}
\includegraphics[angle=0,width=4.5in]{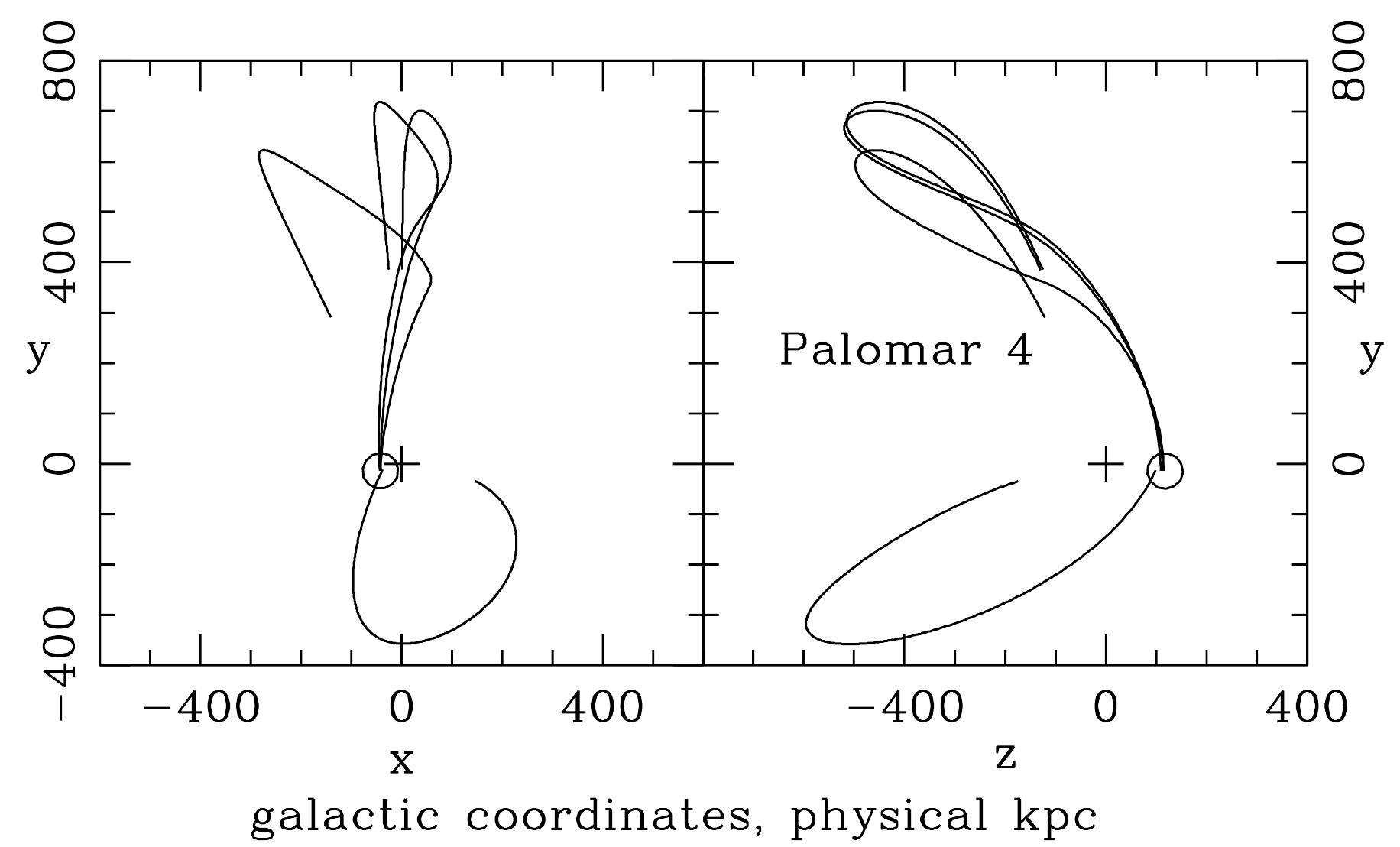} 
\caption{\small Best-fitting solutions for the globular cluster Palomar 4 are plotted as in Figures~\ref{Fig:Crater} and \ref{Fig:AM}.}
\label{Fig:Pal4}
\end{center}
\end{figure}

Acceptable trajectories that fit quiet initial conditions for the third most distant of the  GGCs, Palomar\,4, are shown in Figure~\ref{Fig:Pal4}. The three curves approaching from positive $y$ are different solutions, at different stationary points of the action. The complicated shapes show the effect of the mass exterior to MW, and the differences show different ways to move through the anisotropic external gravitational field. The directions of approach to MW from positive and negative $y$ somewhat resemble the symmetry shown in Figure~\ref{Fig:AMfwd} for AM\,1, and it may be significant that Palomar\,4 also has small galactocentric radial velocity. 

The small radial velocity of Palomar\,4 is consistent with a near circular orbit or a position near perigalacticon or  apogalacticon. Quiet initial conditions are consistent with perigalacticon, but apparently not consistent with a near circular orbit. It seems difficult but maybe not impossible to see how a Palomar\,4 formed in the proto-MW could have been left in a near circular orbit. Zonoozi, Haghi, Kroupa,  K{\"u}pper, and Baumgardt (2017) argue that the stellar distributions and the mass function in Palomar\,4 are best understood if this cluster is near apogalacticon, and has been disturbed by close passages of MW at perigalacticons~$\sim 5$~kpc. The apogalacticon option, with the formation of this cluster in the proto-MW,  might be confirmed by a stream of stars following the last close passage, which would be very interesting. 

\begin{figure}[ht]
\begin{center}
\includegraphics[angle=0,width=4.5in]{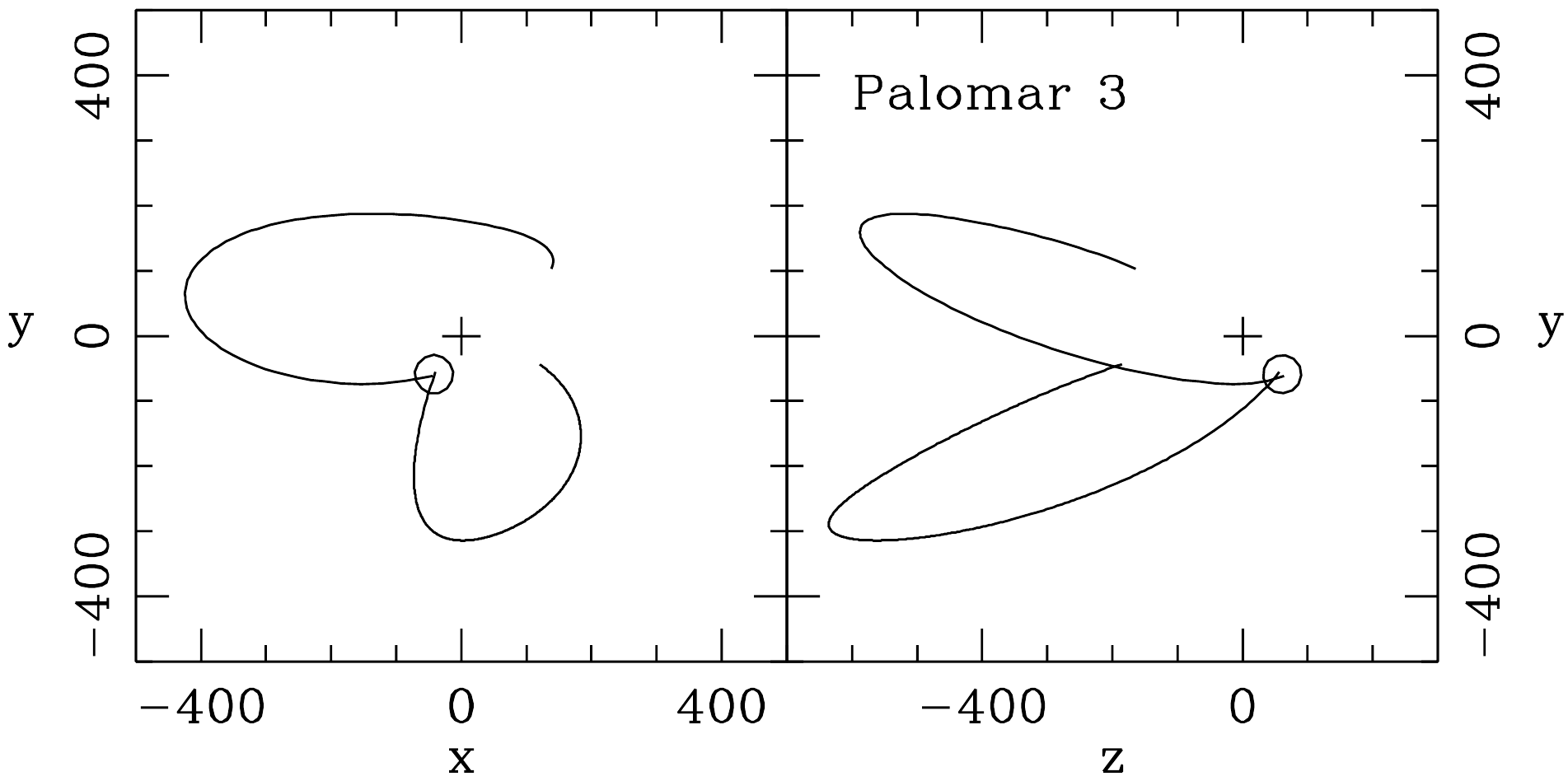} 
\caption{\small The best-fitting solutions for Palomar 3, plotted as in Figures \ref{Fig:AM} to \ref{Fig:Pal4}.}
\label{Fig:Pal3}
\end{center}
\end{figure}

Palomar\,3 has a measured proper motion. The two acceptable fits to the measured position and velocity, within the uncertainties and quiet initial conditions,  are shown in Figure~\ref{Fig:Pal3}. The solution at more negative $y$ has components of proper motion within the measurement uncertainties, redshift low by 10~km~s$^{-1}$, and distance 8 \% below catalog. The other solution better fits the catalog redshift and distance but it has proper motion $1.9~\sigma$ large in the direction of increasing $\alpha$ and $2.4~\sigma$ low toward increasing $\delta$, which is somewhat questionable. In both trajectories this globular cluster is approaching MW for the first time from below the galactic plane, from rather different directions, despite the constraint on the proper motion. The next best-fitting solution, not plotted, has redshift large by 33~km~s$^{-1}$ and proper motion $2.1~\sigma$ low toward increasing $\delta$. 

\begin{figure}[ht]
\begin{center}
\includegraphics[angle=0,width=4.5in]{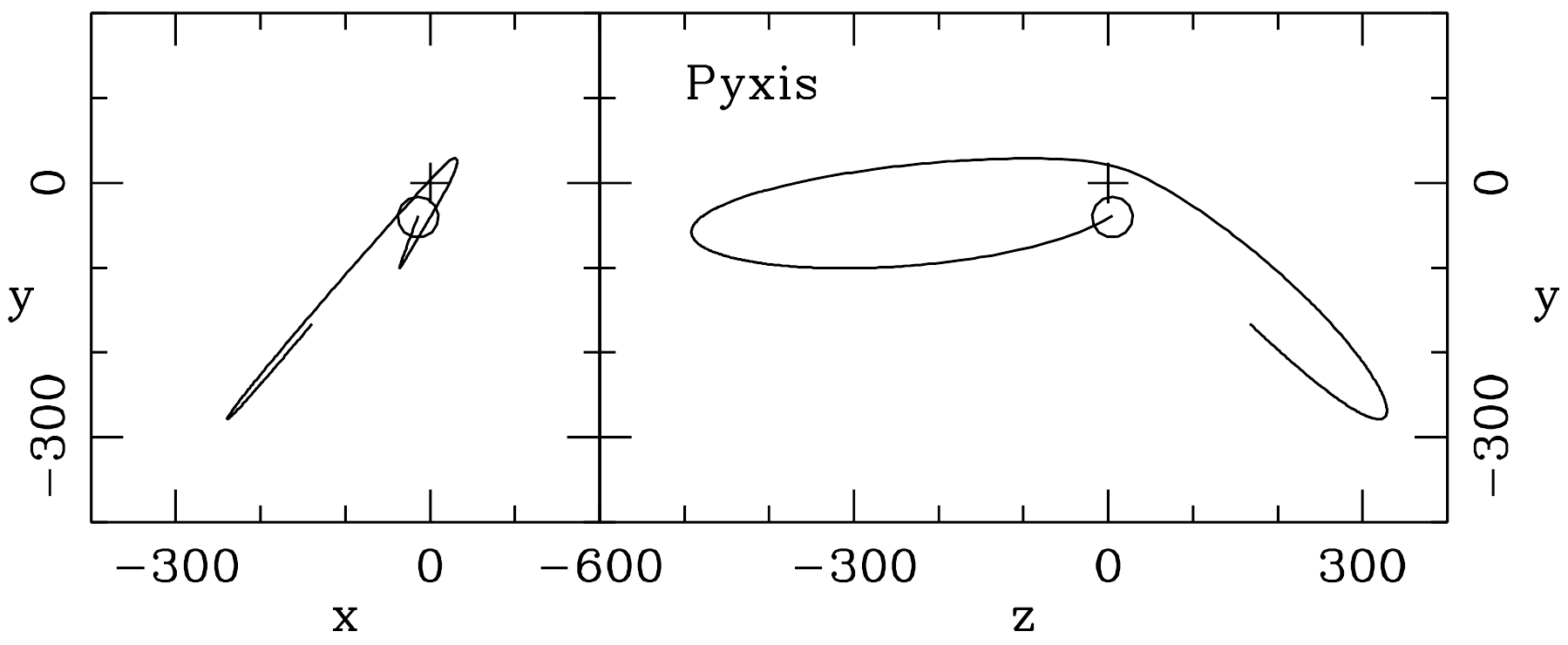} 
\caption{\small Solution for the trajectory of the globular cluster Pyxis, plotted as in Fig.~\ref{Fig:Pal3}.}
\label{Fig:Pyxis}
\end{center}
\end{figure}
Figure~\ref{Fig:Pyxis} shows the best-fitting solution for the globular cluster Pyxis, which also has a measured proper motion. The redshift and distance differ from catalog by about one nominal standard deviation, the proper motion toward increasing $\alpha$ is 2.2 standard deviations above the measurement, and the proper motion to increasing $\delta$ is 1.3 standard deviations larger than the measurement (from Fritz, Linden,  Zivick, et al.\ 2017). In the next best solution, not plotted, the redshift is low by 39~km~s$^{-1}$ and the proper motion components are off by 2.1 and 1.8 standard deviations, which in combination make the solution appear unpromising. The trajectory shown in Figure~\ref{Fig:Pyxis} has Pyxis passing 30 kpc from MW at redshift $z=0.9$. This is close enough that the trajectory is sensitive to the detailed distribution of mass in MW. 

\subsection{Tidal Streams}\label{sec:tails}

Model trajectories of globular clusters might be tested, and degeneracies of solutions reduced, by comparing models and observations of gravitationally-produced tidal streams or tails (Navarrete, Belokurov, and Koposov 2017; Carlberg 2017). This preliminary discussion of the situation for NAM solutions for the outer GGCs is motivated by the important detection of likely-looking tidal tails in Eridanus and Palomar\,15 (Myeong, Jerjen, Mackey, and Da Costa 2017).

\begin{figure}[ht]
\begin{center}
\includegraphics[angle=0,width=6.in]{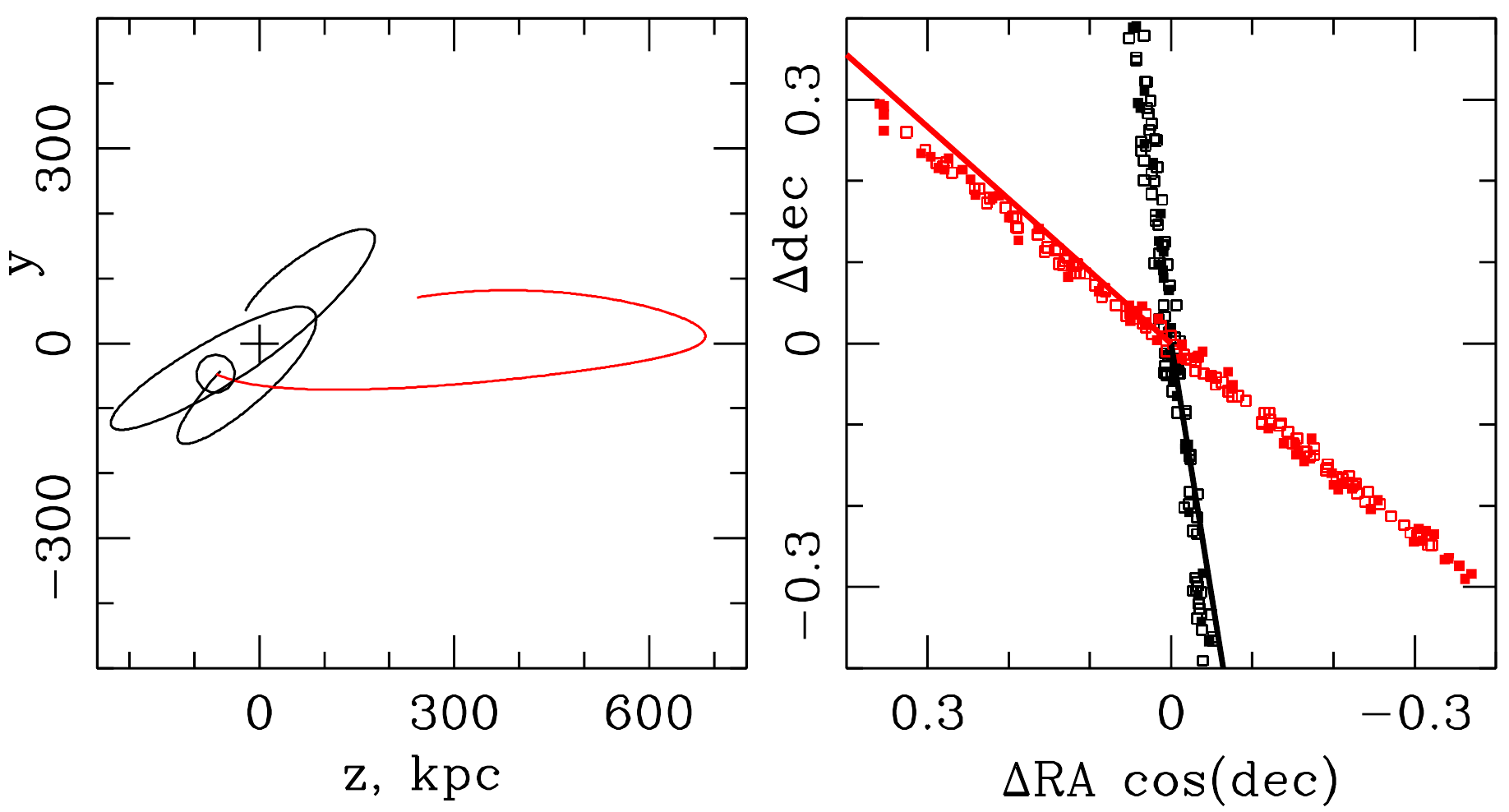} 
\caption{\small Two solutions for the trajectory of the globular cluster Eridanus. In the right-hand panel the trajectories ending at the origin run close to the streams of test particles initially uniformly and isotropically distributed around Eridanus and initially moving with the same coordinate velocity.}
\label{Fig:Eridanus}
\end{center}
\end{figure}

Figure~\ref{Fig:Eridanus} shows two solutions for the path of Eridanus. The more tightly bound trajectory plotted in black fits the catalog redshift and distance within one nominal standard deviation. The path plotted in red has redshift high by 12~km~s$^{-1}$ and distance high by 12\%, perhaps acceptable within the uncertainties of the computation and measurements. The solid curves ending at the origin in the right-hand panel show the approach of Eridanus to its present  position in heliocentric equatorial angular coordinates. The filled squares are the present angular positions of 100 test particles uniformly distributed at random within 10~physical parsecs of Eridanus at redshift $z=9$; the open squares are the present angular positions of 100 test particles uniformly distributed at random within 75~physical parsecs of Eridanus at redshift $z=1$. The initial coordinate velocities of the particles, at $z=9$ or $z=1$, are the same as that of Eridanus, meaning the initial distribution is expanding with the general expansion of the universe. The more tightly bound trajectory in Figure~\ref{Fig:Eridanus} completes close to two orbits around MW. As discussed in Section~\ref{sec:degeneracies} the direction of approach to its present position, and the orientation of its gravitational stream, are not likely to be reliably connected to the initial conditions. The trajectory plotted in red has Eridanus approaching MW for the first time from maximum physical distance $\sim 700$~kpc. The form of this trajectory certainly is sensitive to the mass of MW, and likely M\,31, but within the general parameters of the mass model the direction of approach seems reasonably well based. And it may be significant that the orientation of this more secure stream is close to what is observed (Myeong et al. 2017). 

Myeong et al. also observed apparent tidal tails around the globular cluster Palomar\,15. In the two NAM solutions that best fit the measured redshift and distance the trajectory of this object has completed five close passages of MW before the present. The model solutions for the orientation of the stream thus do not seem trustworthy at the present state of understanding of the mass distribution.

The initially spherical distributions of particles are readily pulled into long streams during the final approach to MW, as we see from the similar streams originating at $z=9$ and $z=1$, because the cluster and particles are massless, moving in the gravitational field of the twelve massive actors.  The simulation should be repeated with a massive model for the globular cluster, and maybe even massive particles pulled from the cluster. 

\section{Discussion}\label{sec:discussion}

The first goal of this analysis was to add to the tests of the model for the mass distribution and evolution in and around the Local Group of galaxies found by fitting measurements of dwarf galaxies (Peebles 2017). Measured and computed redshifts and distances under the postulate of quiet initial conditions agree equally well for the McConnachie  (2012; 2015) sample of 49 dwarf galaxies at distances 50~kpc to 1~Mpc (Peebles 2017 Fig.~2), the 6 McConnachie dwarf galaxies at 30 to 50~kpc added in the present analysis, and the 12 galactic globular clusters at distances greater than 30~kpc (Fig.~\ref{Fig:cz} in this paper). That is, the two sets of objects added to the analysis in Peebles (2017) do not indicate any problem with the  mass model. A fuller test would seek solutions in which the Milky Way is less massive, as may be indicated by the Eadie et al. (2017a,b) analysis. But testing this by using a lower mass for MW is complicated. The outer GGCs and the dwarf galaxies are far enough from MW that the gravitational influence of M\,31 may be important, as one sees from the interestingly curved trajectories shown in Section~\ref{sec:degeneracies}. And it may be recalled that in the quiet picture for the trajectory of the Large Magelanic Cloud a massive object in addition to MW, M\,31, and LMC is required to produce the motion of LMC normal to the MW-M\,31-LMC plane (Peebles 2010). A fuller test of the mass model thus requires reanalysis of the motions of MW and M\,31, with the many dwarf galaxies and GGCs surrounding them, in the gravitational field of a fair sample of the mass in and around the Local Group. This might best await more data from surveys in progress, and reconsiderations of the cataloged data. 

\begin{figure}[ht]
\begin{center}
\includegraphics[angle=0,width=4.5in]{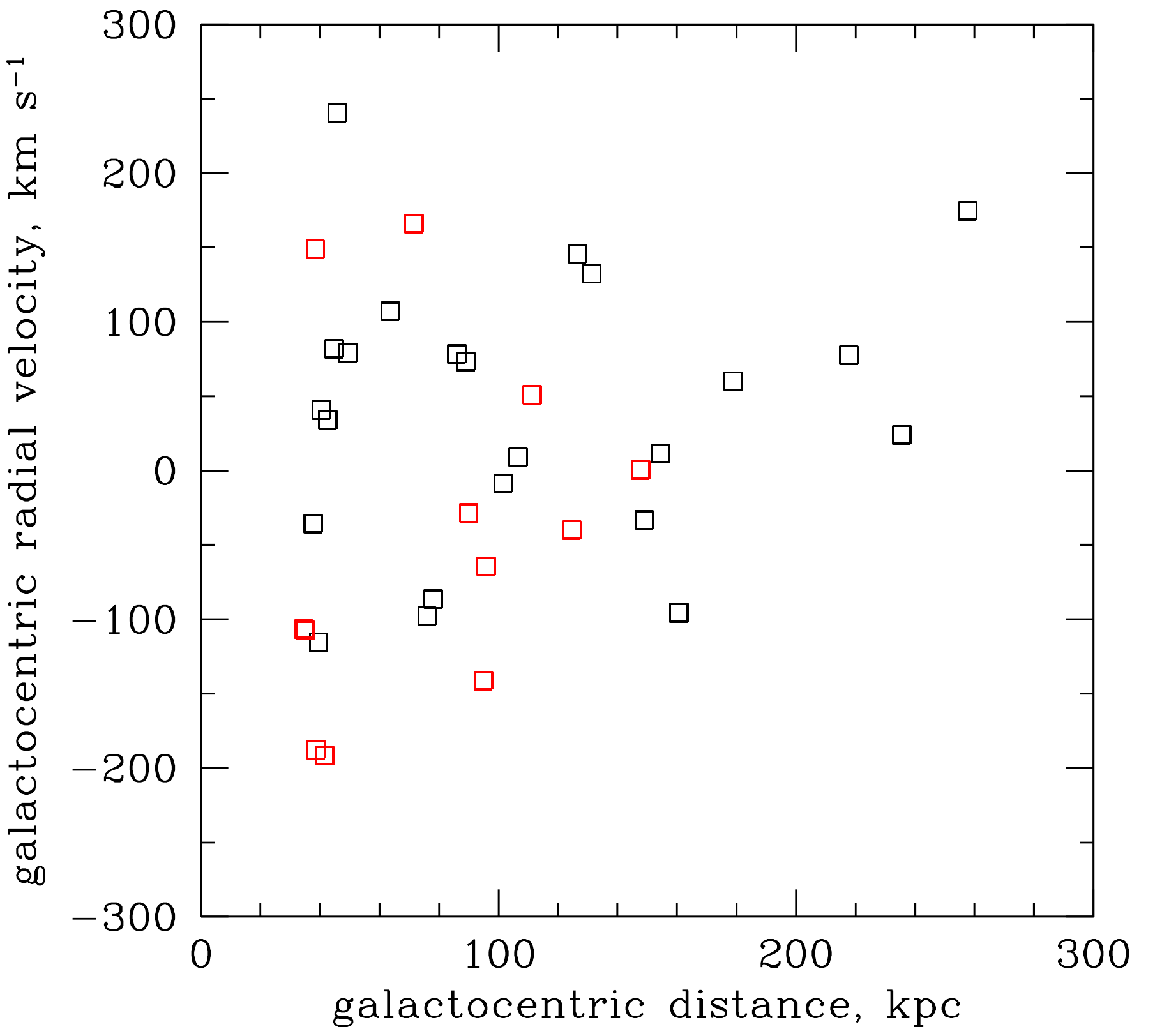} 
\caption{\small Redshifts and distances referred to the center of the Milky Way are plotted in red for globular clusters and black for dwarf galaxies.}
\label{Fig:dzGC}
\end{center}
\end{figure}

The second goal was to add to the evidence about formation of the outer galactic globular star clusters, and the dwarf galaxies at similar distances, whether in the violence of assembly of the proto-Milky Way or other large galaxies or in the quiet tranquility of formation during the onset of nonlinear departures from the initially near homogeneous baryon distribution. The comparison of distributions of galactocentric redshifts and distances of GGCs and dwarf galaxies in Figure~\ref{Fig:dzGC}, with other details in Figure~4 in Laevens, Martin, Sesar,  et al. (2014), offer hints to how the modes of formation might compare. The horizontal axis in the figure is the galactocentric distance in Table~1; the vertical axis is the approximation to the galactocentric radial velocity in Table~1. The GGCs tend to have smaller radial velocities than the dwarfs. But the mean of the galactocentric redshift components of the 17 dwarfs at galactocentric distances between 30 and 150~kpc is $38 \pm 23$~km~s$^{-1}$, and the mean for the 12 GGCs is $-42 \pm 34$~km~s$^{-1}$. Both are consistent with zero mean, about as many going out as coming in, within two standard deviations. 

The count of tabulated GGCs relative to dwarfs in Figure~\ref{Fig:dzGC} decreases sharply beyond 100~kpc, but it is not clear whether this may be attributed to greater incompleteness of the GGC sample. The dynamical model trajectories illustrated in Figures~\ref{Fig:N7006} to~\ref{Fig:Pyxis}  have some outer GGCs approaching MW from much farther than 100~kpc and, unless the impact parameter is small enough to slow or destroy these objects, they may be expected to move out again to similarly large distances. In this picture some GGCs surely have already fallen in and have moved well away from MW again. Thus a reasonably clear prediction of the quiet formation model is that there are substantially more GGCs at distances beyond 300~kpc than the twelve cataloged at distances of  30~kpc to 150~kpc. If GGCs arrived in the halos of dwarf galaxies in trajectories  plunging in the manner of the Magellanic Clouds then, since dwarfs are more likely than their globular clusters to be tidally disrupted or  slowed by dynamical drag, one again expects the ratio of globulars to dwarfs to increase beyond 100~kpc. If there are not considerable numbers of globular clusters beyond 300~kpc it will call for serious reconsideration of the trajectories in Section~\ref{sec:degeneracies}  and ideas about GGC formation. 

If the GGCs formed under the conditions of strong gravitational acceleration and dissipation within parts of massive proto-galaxies then it ought not to be consistently possible to fit the measured redshifts, distances, and proper motions to solutions to the equations of motion computed under quiet initial conditions. There are problems in fitting model to data: P17 does not yield satisfactory fits to the eight dwarfs in the NGC\,3109 and DDO\,210 associations, which have systematically large model distances and/or low model redshifts. This problem, perhaps with aspects of the mass model, remains open. But there are no serious problems in P17 with the 49 dwarfs closer than 1~Mpc, and there are no problems in fitting  to primeval conditions the data for the six closer dwarfs and twelve outer GGCs considered here. This consistency of acceptable fits argues for formation in primeval conditions reasonably well represented by the adopted mass model. 

There are the complications illustrated in Figures~\ref{Fig:N7006} to~\ref{Fig:Pyxis}. I was not able to find a trajectory for NGC\,7006 (Fig.~\ref{Fig:N7006}) consistent with the measured redshift, distance, and proper motion that has this globular cluster falling toward MW from a distance large enough that one might trust the initial conditions. The evidence certainly allows formation of NGC\,7006 under near primeval conditions and close enough that it has orbited MW several times. But the evidence equally well allows formation of NGC\,7006 in some major part of the proto-MW, provided the position and velocity at formation allow this globular cluster to occupy the outer parts of the assembled Milky Way. The solutions for other GGS allow them to have completed several passages of MW, obliterating details of initial conditions, or else to be approaching MW for the first time from reasonably secure initial  conditions within the mass model. The only reasonably acceptable solutions I have found for the most distant GGCs in the sample, Crater/Laevens\,I, AM\,1, and Palomar\,4,  have these objects falling into MW for the first time, but there is serious degeneracy of solutions. Proper motion measurements reduce the degeneracies and test trajectories from quiet initial conditions, though an unambiguous result is not guaranteed. Thus the measured proper motion of Palomar\,3 allows the two quite different trajectories in Figure~\ref{Fig:Pal3}. To be considered also is that I may  not have found trajectories for the outermost of the catalogued GGCs that have circled MW because I have assumed the wrong initial conditions. This can be investigated for clusters that have a proper motion measurement by finding the frequency of occurrence of orbiting trajectories in a sample that traces back in time from present conditions that fairly represent the distance, redshift, and proper motion within the  measurement uncertainties. 

The discussion of tidal tails in Section~\ref{sec:tails} is a preliminary step that suggests the orientations are a powerful discriminant, if deep images detect tails attached to significant numbers of outer globular clusters and dwarf galaxies. The constraints on dynamics in and around the Local Group may be further improved by proper motion measurements. Surveys in progress may be expected to discover more dwarf galaxies and maybe globular clusters at distances from a few tens of kiloparsecs to a few megaparsecs, adding to the critical data for dynamics: measured redshifts and distances. Less exciting but equally important are critical data for the catalogued globular clusters. A few redshifts seem to be seriously uncertain. Are all the catalogued globular cluster distances really good to 5\%?

\acknowledgments
I have profited from advice from Gwen Eadie,  Bill Harris, Alan McConnachie, Ed Olszewski, Ed Shaya, Brent Tully, and Karina Voggel.

\end{document}